\documentclass[aps,pre,preprint,amsmath,amssymb,superscriptaddress,showpacs,floatfix]{revtex4-1}
\usepackage{graphicx}
\usepackage{dcolumn}
\usepackage{bm}
\usepackage[colorlinks, allcolors=blue]{hyperref}
\usepackage[usenames, dvipsnames]{color}
\usepackage{setspace}
\usepackage{scalerel}
\usepackage{stackengine}
\usepackage{comment}
\usepackage{csquotes}
\stackMath
\newsavebox\tmpbox

\begin{document}

\title{Quantum Phase Transitions in Skewed Ladder Systems}

\author{Sambunath Das}
\affiliation{Institute of Physics (FZU), Czech Academy of Sciences, Na Slovance
2, 182 00 Prague, Czech Republic}

\author{Dayasindhu Dey}
\affiliation{UGC-DAE Consortium for Scientific Research, Indore - 452001, Madhya
Pradesh, India}

\author{Rajamani Raghunathan}
\affiliation{UGC-DAE Consortium for Scientific Research, Indore - 452001, Madhya
Pradesh, India}

\author{Zolt\'an G. Soos}
\affiliation{Department of Chemistry, Princeton University, Princeton, New Jersey 08544, USA}

\author{Manoranjan Kumar}
\email{manoranjan.kumar@bose.res.in}
\affiliation{S. N. Bose National Centre for Basic Sciences, Block - JD, Sector -
III, Salt Lake, Kolkata - 700106, India}

\author{S. Ramasesha}
\email{ramasesh@iisc.ac.in}
\affiliation{Solid State and Structural Chemistry Unit, Indian Institute of Science, Bengaluru 560012, India}
\date{\today}
\begin{abstract}
In this brief review, we introduce a new spin  ladder system called skewed spin ladders and discuss the exotic quantum phases of this system. The spin ladders studied are the 5/7, 3/4 and 3/5 systems corresponding to alternately fused 5 and 7 membered rings; 3 and 4 membered rings; and 3 and 5 membered rings. These ladders show completely different behaviour as the Hamiltonian model parameter is changed. When the Hamiltonian parameter is increased the 5/7 ladder switches from an initial singlet ground state to progressively  higher spin ground state and then to a reentrant singlet state before finally settling to the highest spin ground state whose spin equals the number of unit cells in the system. The 3/4 ladder goes from a singlet ground state to a high spin ground state with each unit cell contributing spin 1 to the state, as the model parameter is increased. The 3/5 ladder shows a singlet ground state for small parameters and high spin ground state for intermediate values of the parameter and for still higher parameters, a reentrant singlet ground state. They can also show interesting magnetization plateaus as illustrated by studies on a specific spin ladder.
\end{abstract}
\maketitle
\section{\label{sec:intro}Introduction}
Quantum phase transitions occur in the ground state (GS) of a model Hamiltonian with noncommuting terms and correspond to a qualitative change in the GS of the system when the parameters of the model are varied. While finite temperature thermodynamic transitions are driven by a competition between entropy and enthalpy, the quantum phase transitions are driven by quantum fluctuations due to the non-commuting parts of the Hamiltonian, leading to a change in the ground state of the system. An example is the change in GS of a one-dimensional extended Hubbard model from a spin density wave state to a bond order wave state or from a bond order wave state to a charge density wave state~\cite{Hirsch1,Hirsch2,Nakamura_2000,Sengupta,kumar,Klein_2017}. Best known model Hamiltonians exhibiting quantum phase transitions are the one-dimensional Hubbard~\cite{hubbard63,hubbard64} and long range Hubbard models (also known as the Pariser-Parr-Pople models)~\cite{pariser_parr53,pople53} and the Heisenberg spin chains with competing interactions~\cite{ckm69a, ckm69b, hamada88, chubukov91, chitra95, white96, itoi2001, mahdavifar2008, sirker2010, mk2015, soos-jpcm-2016, mk_bow, chubukov91, mk2012, mk2015, vekua2007, hikihara2008, sudan2009, dmitriev2008, meisner2006, meisner2007, meisner2009, aslam_magnon, kecke2007}. The Hubbard model was invented to study itinerant electron magnetism and has been the cornerstone for studying electron correlations in metallic systems. In its simplest form, it assumes one valence orbital per site and a nearest neighbor hopping integral and electron repulsion only when two electrons occupy the same orbital~\cite{hubbard63,hubbard64,hubbard64b,Kemeny_rev,Rado1966}. This model is extensively studied in various geometries~\cite{Arovas,Lieb_EH}. The PPP model was invented to study correlated conjugated molecular systems wherein the electron repulsions are long ranged. Heisenberg models result in the limit of very strong electron-electron interactions in the correlated electron models where the charge degrees of freedom are frozen and only spin fluctuations are at play. Heisenberg models are important in the study of one dimensional~\cite{hase2004, masuda2004, mizuno98, riera95, kamieniarz2002,drechsler2007, dutton2012, dd2018, sirker2010}, quasi-one dimensional~\cite{Dagotto,Nagata,Windt,Notbohm} and higher dimensions~\cite{Claudine_2011,Chaloupka,Lee,Gong} magnetic systems. Interestingly these simple models are sufficient to model many of the electronic properties of materials~\cite{Mcqueen_2009,Kumar_2011,Sarkar_2021}. The spin chain with uniform antiferromagnetic exchange constants between nearest neighbor (NN) spins was exactly solved by Bethe using an ansatz which goes by his name~\cite{bethe31,hulthen38}. In the limit of infinite chain length (thermodynamic limit) the singlet GS is gapless and the triplet state is lowest excitation~\cite{Lieb_1961}. The dimerization of the spin chain, characterized by alternately strong and weak NN exchange interactions along the chain occurs, opens a gap between the singlet and triplet state \cite{Johnston_2000,chitra95,Sudip_2020,Ren_2006,Ding_2009,Klein_2021}.
Experimentally more interesting one dimensional spin models are the dimerized spin chains with second neighbor exchange interactions which frustrate the NN interactions~\cite{Hase_1993,Miljak_2005,Ren_2007,Jacobs_1976,hybrid}.
\begin{figure}[h]
  \includegraphics[width=3.0 in]{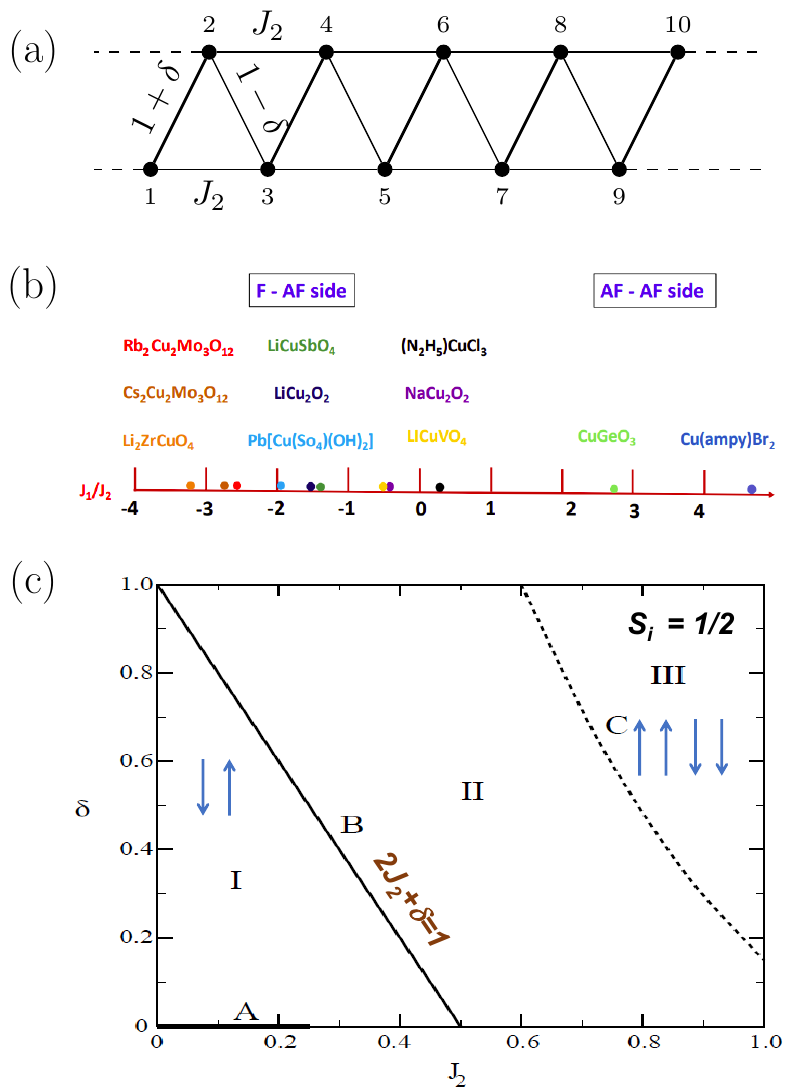}
	\caption{\label{fig:j1j2}(a) The dimerized and frustrated spin chain. The average NN exchange constant is set to 1 and defines the energy scale. $J_2$ is the next nearest neighbor (NNN) exchange constant and $\delta$ is the dimerization parameter. (b) Quasi 1-d systems described by the model and the ratio of the NN to NNN exchange constants of the system. The NNN exchange is always antiferromagnetic. (c) Phase diagram of the model. The dark line A on the $J_2$ axis is the gapless line. Along line B, the model is exactly solvable and at $\delta=0$ and $J_2 = 0.5$, the Kekul\'e state with NN singlets is the GS. Region I is where the GS is gapped and the GS can be described by a N\'eel phase. Region II corresponds to spin structure factor peaking between $\pi/2$ and $\pi$. Region III is the region with double period N\'eel phase (reproduced from ref.~\cite{chitra95} and ~\cite{Pati1997} with permissison from Americal Physical Society (APS) and Institute of Physics (IOP) Publishing, copyright 2023).}
\end{figure}

The dimerized and frustrated spin chain can be pictorially described as in Fig.~\ref{fig:j1j2}(a), where the NN exchanges alternate between $1+\delta$ and $1-\delta$ and the second neighbor exchange is $J_2$. Fig.~\ref{fig:j1j2}(b) shows the materials described by the model and the corresponding $J_2/J_1$ values~\cite{hase2004, masuda2004, mizuno98, riera95, kamieniarz2002, drechsler2007, dutton2012, dd2018, sirker2010}. When $\delta=0$, the model shows a gapless GS in the range $0 < J_2/J_1 < 0.2411$ and gapped phase in all other regions of the parameter space~(Fig.\ref{fig:j1j2}(c)). The spin structure of the GS in region I is N\'eel type with the structure factor $S(q)$ peaking at $q = \pi$, in region II the peak in $S(q)$ gradually decreases from $\pi$ until it hits region III where the maxima in $S(q)$ is at $\pi / 2$ corresponding to a double period N\'eel phase~\cite{chitra95,mk2007}. On the line $2 J_2 + \delta = 1$, one of the two Kekul\'e states corresponding to singlet between spins on sites $2j$ and $(2j+1)$ or between $(2j-1)$ and $2j$ is the GS. At $\delta = 0$ and $J_2 = 0.5$, the two Kekul\'e states are degenerate and this point is known as the Majumdar-Ghosh point~\cite{ckm69a,ckm69b}. In the large $J_2/J_1$ limit the $J_1-J_2$ chain can be mapped on to a zigzag ladder whose rung and leg exchanges are $J_1$ and $J_2$ respectively~\cite{chitra95}.

Regular two and three leg spin ladder systems are experimentally known for many decades. Initial interest in spin ladders was to obtain superconductivity in these systems by hole doping and it was conjectured that it could be observed in (VO)$_{2}$P$_{2}$O$_{7}$ system~\cite{Dagotto}. It was indeed observed by Nagata et al in the two leg ladder compound Sr$_{14-x}$Ca$_{x}$Cu$_{24}$O$_{41}$ under pressure~\cite{Nagata}. Recently there has been considerable interest in spin ladders to
find bound magnon states~\cite{Windt,Notbohm}. Current interest in spin ladders centres around magnetization plateaus and their ground states~\cite{Sasaki,Vlady_2011,meisner2007,tandon99,okunishi_jpsj2003, okunishi_prb2003}. Spin ladder systems are of interest as they bridge the gap between the one and two dimensional systems. Ladders with even number of legs and odd number of legs behave qualitatively differently. The former exhibits a spin gap while akin to the integer spin chains while the latter are gapless like the half odd-integer spin chains~\cite{haldane83a,haldane83b}. The odd leg ladders also exhibit magnetization plateaus and conform to the Oshikawa, Yamanaka and Affleck (OYA) criterion which follows from the existence of a spin gap between the lowest energy states in the $S^z$ and $S^z +1$ sectors in the thermodynamic limit~\cite{oya97}. The magnetic plateau in magnetization $M$ and magnetic field $B$ curve represents a stable ground state at a certain critical field $B$. The $M-B$ curve of $J_1-J_2$ model on a zigzag ladder also shows a $m=1/3$ $(m= M/M_s)$ plateau phase for $|J_2/J_1| > 0.6$ for both antiferromagnetic~\cite{okunishi_jpsj2003}, and ferromagnetic $J_1$ and antiferromagnetic $J_2$. The plateau feature in $M-B$ curve is common in many frustrated systems like Cu$_3$(CO$_3$)$_2$(OH)$_2$~\cite{kikuchi2005,kikuchi2006,gu2006}, Ca$_3$Co$_2$O$_6$~\cite{zhao2010,maignan2004,hardy2004}, Sr$_3$Co$_2$O$_6$~\cite{wang2011}, Sr$_3$HoCrO$_6$~\cite{hardy2006}, SrCo$_6$O$_{11}$~\cite{ishiwata2005} and CoV$_2$O$_6$~\cite{yao2012,lenertz2011,he2009}; they all have a plateau at $m =$ 1/3. The frustrated ladder compound NH$_4$CuCl$_3$ shows two plateaus at $m = 1/4$ and 3/4~\cite{shiramura98}.

\begin{figure}
  \includegraphics[width=3.3in]{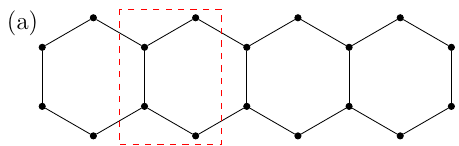}
  \includegraphics[width=3.3in]{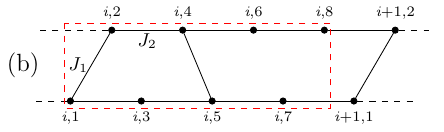}
  \includegraphics[width=3.3in]{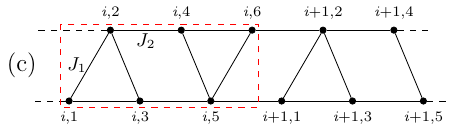}
  \includegraphics[width=3.3in]{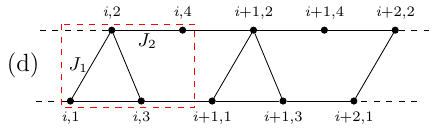}
        \caption{\label{fig:schematic}Schematic diagrams of (a) an oligoacene (b) 5/7 skewed ladder (c) 3/4 skewed ladder and (d) 3/5 skewed ladder. The dashed rectangle in each diagram depicts a typical unit cell for the system. The letter `$i$' represents the index of the unit cell, and the numbers 1, 2,... represent the numbering of the spins within the unit cell. The solid line indicates antiferromagnetic exchange interaction between corresponding sites.}
\end{figure}

The zigzag spin ladder with alternately missing rungs is structurally like the oligoacenes (Fig.~\ref{fig:schematic}(a)) and fermionic models of this system have been studied extensively~\cite{thomas2012} and these systems look like skewed ladders. In this review we are interested in these skewed ladder systems. In these systems some of the rung bonds could be either missing or are not between corresponding sites in the two legs. The skewed ladders resemble catacondensed ring systems, for example the ladder shown in Fig.~\ref{fig:schematic}(b) corresponds to fused azulenes, consisting of alternately five and seven membered rings and hence are called 5/7 skewed ladders. The short oligomers of fused azulene (up to six azulene units) were studied using unrestricted DFT and spin models~\cite{Qu}, which showed a triplet ground state for longer oligomers. Recently Rano et al have studied a related system, namely the fused acene-azulene systems using ab initio methods, in the quest for triplet ground states~\cite{Rano}. There are many theoretical attempts to design magnetic ground state of hydrocarbon based systems similar to skewed ladder ~\cite{Valentim_2020,Valentim_2022,Rano,Chiappe_2015}. The 3/4 and 3/5 skewed ladders are shown in Figs.~\ref{fig:schematic}(c) and (d) respectively. The skewed ladder systems are inherently frustrated models, in that the spin arrangement in the system cannot be up and down on all the bonds as required by antiferromagentic exchange interaction for minimum energy. In the Fermionic context, they do not have charge conjugation or electron-hole symmetry and frustration leads to increased kinetic energy and increased GS energy, compared to an equivalent unfrustrated model. In such Fermionic models, the spin gap is expected to reduce while the optical gap, which is largely controlled by electron correlations, is expected to show only small shifts. This, in principle, could help in designing systems which are suitable for singlet fission in the context of organic photovoltaics. The 5/7 skewed ladder may be found at the grain boundaries in graphene~\cite{Huang2011,Kochat,Balasubramanian2019}. 3/4 ladder can be mapped to coupled trimer spin-1/2 systems in azurites~\cite{Kang_azurite,Honecker_azurite,Rule_azurite}.

In this article, we will review some of our recent results on skewed spin ladders~\cite{geet,plateau57,Das}. In the next section, we introduce the Hamiltonian of the skewed ladders and the computational methods. In section ~\ref{sec:model1}, we briefly discuss the exact diagonalization and density matrix renormalization group techniques which are the numerical methods employed in this study. In section~\ref{sec:gs}, we discuss our results on the low energy states of some skewed ladders. In section~\ref{sec:plateau}, we discuss the magnetization plateaus for 5/7 skewed ladder system. We present a summary and outlook in section~\ref{sec:summary}.
\section{\label{sec:model}Model Hamiltonians}
In this study we focus on isotropic spin-1/2 systems with antiferromagnetic exchange interactions between bonds indicated  in Fig.~\ref{fig:schematic}. Each bond contributes a term $J (\hat{S}_i \cdot \hat{S_j})$ to the Hamiltonian where $J$ is the exchange constant for the bond. The model Hamiltonian of the 5/7 skewed ladder can be written as
\begin{eqnarray}
        H_{5/7} &=& J_1 \sum_i \left(\vec{S}_{i,1} \cdot \vec{S}_{i,2} + \vec{S}_{i,4}
        \cdot \vec{S}_{i,5} \right)  \nonumber \\
        & & +J_2 \sum_i \bigg( \vec{S}_{i,7} \cdot
        \vec{S}_{i+1,1}
        + \vec{S}_{i,8} \cdot \vec{S}_{i+1,2}
        + \sum_{k=1}^6
        \vec{S}_{i,k} \cdot \vec{S}_{i,k+2} \bigg).
\label{eq:ham57}
\end{eqnarray}
where $i$ labels the unit cell and $k$ the spins within the unit cell. Similarly, the model Hamiltonian for the 3/4 and 3/5 systems can be written as
\begin{eqnarray}
        H_{3/4} &=& J_1 \sum_i \left[\left(\vec{S}_{i,1} + \vec{S}_{i,3}\right)\cdot \vec{S}_{i,2} +
        \left(\vec{S}_{i,4} + \vec{S}_{i,6} \right) \cdot \vec{S}_{i,5} \right]  \nonumber \\
        & &+J_2 \sum_i \bigg( \vec{S}_{i,5} \cdot
        \vec{S}_{i+1,1}
        + \vec{S}_{i,6} \cdot \vec{S}_{i+1,2}
        + \sum_{k=1}^4
        \vec{S}_{i,k} \cdot \vec{S}_{i,k+2} \bigg),
\label{eq:ham34}
\end{eqnarray}
and
\begin{eqnarray}
        H_{3/5} &=& J_1 \sum_i \left(\vec{S}_{i,1} \cdot \vec{S}_{i,2}
        + \vec{S}_{i,2}\cdot \vec{S}_{i,3}\right) \nonumber \\
        & & +J_2 \sum_i \bigg( \vec{S}_{i,3} \cdot \vec{S}_{i+1,1}
        + \vec{S}_{i,4} \cdot \vec{S}_{i+1,2}
        + \sum_{k=1}^2 \vec{S}_{i,k} \cdot \vec{S}_{i,k+2} \bigg).
\label{eq:ham35}
\end{eqnarray}
To define the energy scale, we have fixed $J_2$ to unity and varied $J_1$. Hence it is a single parameter model ($J_1/J_2$) for a given skewed ladder and all the results quoted are over a range of this parameter. For small system sizes with up to 24 spins, we have employed exact diagonalization scheme with basis states defined by the projection of individual spins along the z-axis. This corresponds to exploiting $S^z$ conservation of the Hamiltonian. For larger system sizes we have employed the density matrix renormalization group (DMRG) method to obtain the lowest few energy states in each $S^z$ sector. The GS spin, $S$, is the lowest value of $S^z$  which  satisfies $(E_0 (S^z=0) - E_0 (S^z+1)) > 0$ and follows from simple spin algebra which demands ($2S+1$) fold degeneracy of a state with spin $S$, corresponding $S^z$ values $-S \leq S^{z} \leq +S $. If there is more than one state with same $S^z$ that satisfies this criterion then the GS is degenerate state with same spin $S$. The spin bond orders, spin-spin correlation functions as well as the spin densities of high spin GS are obtained using the usual methods. For smaller system sizes, both the spin inversion symmetry and the $C_2$ symmetry are exploited to identify the parameters at which the spin density wave or bond order wave sets in.
\section{\label{sec:model1}Numerical Methods}
In the exact diagonalization technique, for the given finite system, we choose the full space of all the states in a chosen $S^z$ sector which is usually a few million states for the largest system size. The basis states are represented as integers with each bit of the integer representing a site and the state of the bit representing the direction of the z-component of the spin. We set up the Hamiltonian matrix in this basis; the resulting matrix is very sparse and we employ the Davidson’s algorithm to obtain a few low-lying states.

In the DMRG method, usually, we start with a system of a few spins, say 2$k$ spins and construct the density matrix of the left (right) half of the system in a given state. If the desired state of the system can be written as a direct product in the basis of the Fock space on the left (right) half of the system as
\begin{eqnarray}
        |\psi\rangle = \sum_{L,R} \psi_{L,R} |L\rangle |R\rangle,
\end{eqnarray}
then the matrix element of the left half density matrix is given by tracing out the right half states as
\begin{eqnarray}
        {\rho}_{{L}{{L}'}} = \sum_{R} \psi_{L,R} \psi_{{L}',R}.
\end{eqnarray}
We diagonalize the left (right) density matrix and use a small number ‘$\chi$’, usually about a thousand, of the dominant density matrix eigenvectors (DMEV) to span the large \textit{`$L$' $(L=2^n)$} dimensional Fock space. The renormalization group (RG) transformation is effected by the matrix $\textbf{O}_{L/R}$ whose columns are the $\chi$ eigenvectors of the left (L) or right (R) density matrix and has $\chi$ columns and \textit{L} rows. This is followed by setting up the Hamiltonian and other spin operator matrices of the left (right) block in the Fock space basis and renormalizing these \textit{L$\times$L} matrices by transforming to the basis of the block states to obtain the $\chi$$\times$$\chi$ renormalized matrices by the relation
\begin{eqnarray}
        T_R =  O^+ T O,
\end{eqnarray}
where the matrix $T$ is the \textit{L$\times$L} matrix of the desired operator and $T_R$ is the renormalized $\chi$$\times$$\chi$ matrix. We now augment the system by adding two new sites (spins) in the middle of the 2$k$ system and obtain the Hamiltonian matrix of the augmented system in the product basis of the DMEV or the block states of the right and left blocks and the Fock states of the new spins. The augmented Hamiltonian matrix is diagonalized and from the desired eigenvector the density matrices of the new left and right half of the system, each with $k$+1 sites is constructed. From this the new RG transformation matrix is obtained and the desired operators of the left and right blocks are renormalized to obtain the RG transformed operators of the $(k+1)$ site left and right blocks. We then augment the system by adding two new sites and as before set up the Hamiltonian matrix of the augmented (2$k$+4) site system and continue the iteration. At every desired system size, we compute the properties such as spin densities, spin correlations and bond orders from the desired eigenstates for that system size.

The procedure outlined above is known as the infinite DMRG algorithm. To improve the accuracy, we have carried out what is known as the finite DMRG algorithm in which at a chosen system size, 2N, the block on one side, say left, is expanded to N and the block on the right is contracted to N-2 and eigenstates of the 2N system are obtained in the basis of DMEVs of the left block with N sites, DMEVs of the right block with N-2 sites and Fock states of two remaining spins. The augmentation and truncation of the left and right blocks are continued until the right block has only one site and the left has (2N-3) sites. Now the reverse sweep is continued until the left block has only one site and the right has (2N-3) sites. This process is repeated until we reach equal sizes for the left and right blocks. We would then have completed one finite DMRG iteration. We can obtain all the desired expectation values from the eigenstates at this stage, which would be an improvement over the previous one. The purpose behind these sweeps is to ensure that the density matrices of the subsystem correspond to those of the target system; in the infinite algorithm they are the density matrices of the intermediate system. Usually five to six finite iterations are required for convergence of the finite DMRG procedure. We have employed periodic boundary condition for exact diagonalization studies and open boundary conditions for DMRG studies. The spin correlation functions decay very rapidly and hence the boundary conditions are inconsequential for the system sizes we have studied using either of the numerical techniques.

The entanglement entropy of a state can easily be calculated from the eigenvalues of the density matrix and is given by
\begin{eqnarray}
        S = -\sum_{i} \gamma_{i} \log_{2} \gamma_{i},
\end{eqnarray}
where S is the von Neumann entanglement entropy and $\gamma_{i}{'}$s are the eigenvalues of the density matrix. Discontinuity in the entanglement entropy of the ground state indicates a quantum phase transition. A more sensitive indicator of a quantum phase transition is the fidelity parameter. The fidelity of a state when a parameter $J$ of the Hamiltonian is varied is given by
\begin{eqnarray}
        F(J) = \lim_{\delta J \to 0} \langle \phi_{g}(J)|\phi_{g}(J+\delta J) \rangle.
\end{eqnarray}
Where $|\phi_{g}(J)\rangle$ is the ground state for the Hamiltonian parameter $J$. When the ground state changes across the Hamiltonian parameter $J_c$, fidelity shows a sharp change as the two ground states across $J_c$ are orthogonal. We find that $J_{c}$ is accurately determined by the fidelity parameter.
\section{\label{sec:gs}Ground State Properties of the 5/7, 3/4, and 3/5 ladders}
Here we discuss the GS properties of the spin-1/2 hetero skewed ladders characterized by neighboring rings having unequal number of vertices. The homogeneous skewed ladders have a singlet GS for all values of the parameter $J_1/J_2$ although they exhibit spin density or bond order wave phases. The skewed ladder systems discussed below are the 5/7, 3/4 and 3/5 skewed ladders.
\subsection{5/7 Skewed ladders}
\begin{figure}
  \includegraphics[width=3.0in]{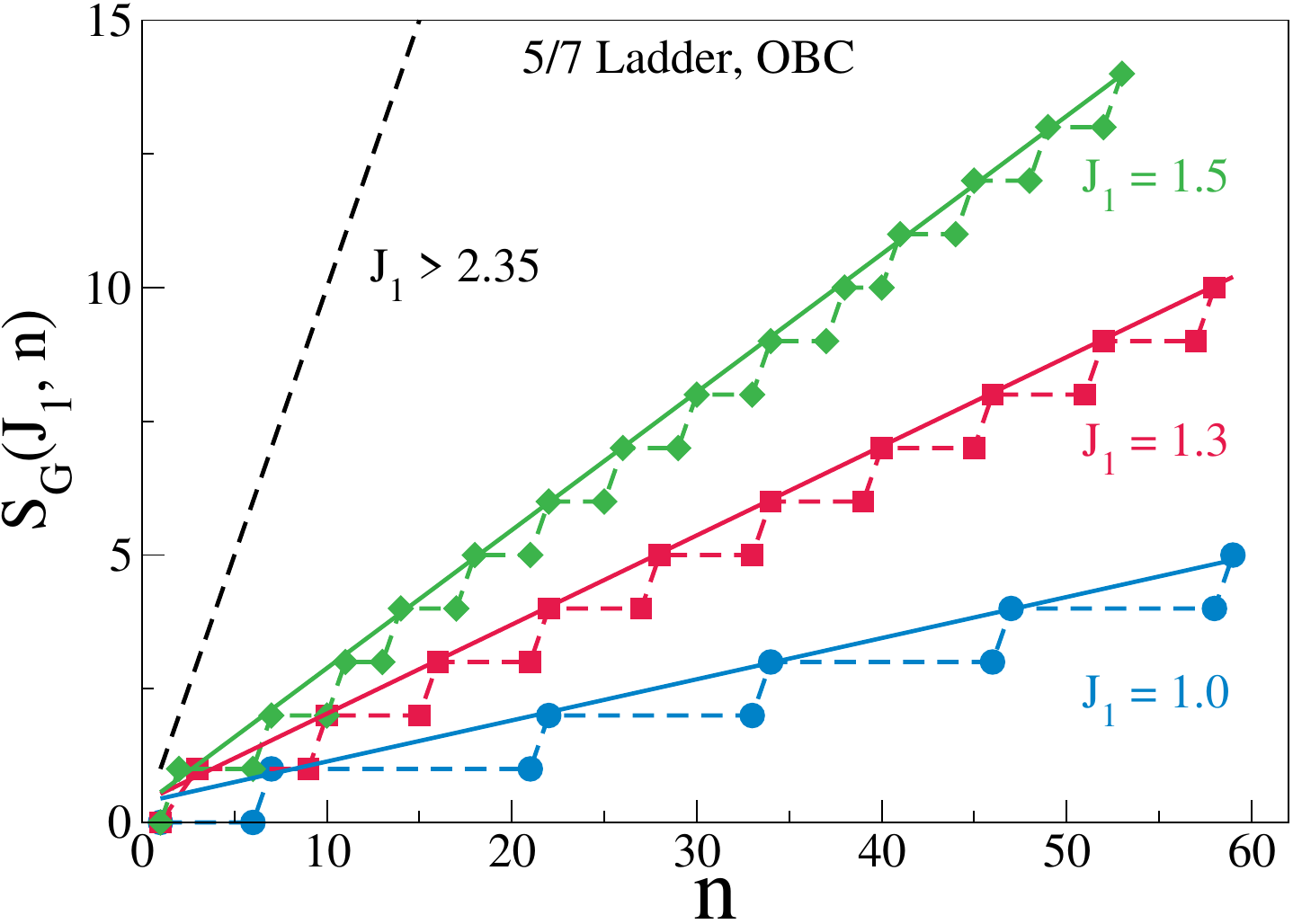}
	\caption{\label{fig_57_res1}The system size dependence of the GS spin $S_{G}(J_{1},n)$ in 5/7 ladders together with the indicated $J_1$. The dashed line is for $J_{1} >2.35$. Reproduced from ref.~\cite{geet} with permission from APS, copyright 2023.}
\end{figure}

\begin{figure}[h]
  \includegraphics[width=3.0in]{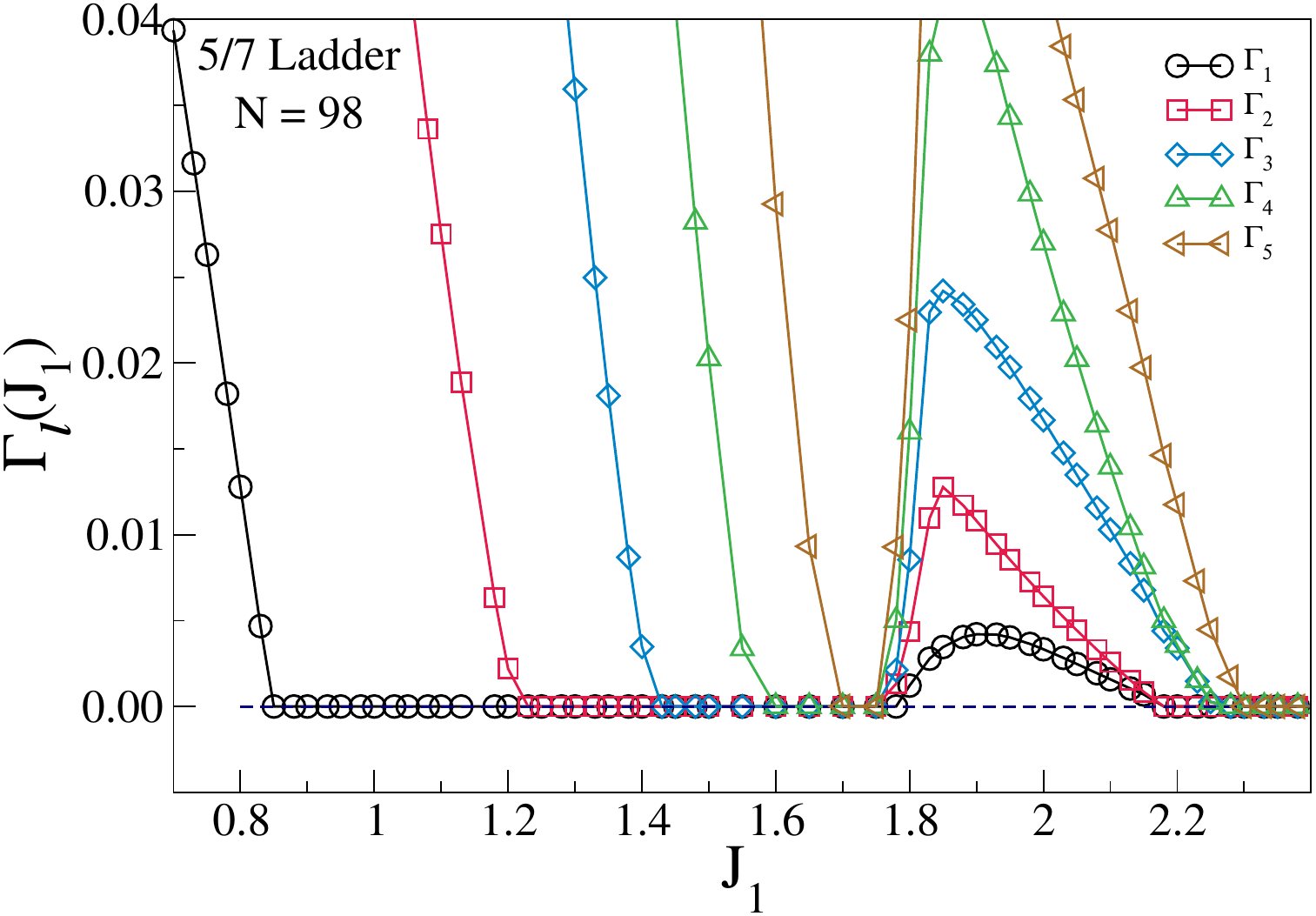}
	\caption{\label{fig_57_res2}The GS spin $S_G(J_1, N)$ of a 5/7 ladder of 98 spins with OBC as a function of $J_1$. The system reenters $S_G(J_1, N)=0$ in the region $(1.75 < J_1 < 2.18)$. Reproduced from ref.~\cite{geet} with permission from APS, copyright 2023.}
\end{figure}

The spin $S_G$ of the GS of this system depends on both the system size and the relative strength of the rung exchange $J_1$ (which is always in units of $J_2$). The GS is a singlet $(S_G = 0)$ for small system sizes, for $J_1 = 1$ and $S_G$ increases gradually with increase in system size. The rate of increase in $S_G$ increases as $J_1$ increases and the rate saturates at $\Delta S_G = 1$ for $\Delta n = 1$ for $J_1 > 2.35$, where $n$ defines the number of unit cells and the system size, $N$, for open boundary condition is given by $N=8n+2$ (Fig.~\ref{fig_57_res1}). In this limit, each unit cell contributes two unpaired spins to the GS spin of the system. The 5/7 ladder also shows exotic behavior in the dependence of $S_G$ as a function of $J_1$ (Fig.~\ref{fig_57_res2}). For a fixed system size, while $S_G$ increases with $J_1$ till $J_1 = 1.75$, the system reenters the singlet GS for $1.75 < J_1 < 2.18$. Beyond $J_1 = 2.18$ the system rapidly regains higher $S_G$ values and attains the highest GS spin of $S_G = n$. In the re-entrant phase $(1.75 < J_1 < 2.18)$, the spin correlations indicate a long wavelength spin density wave.
\begin{figure}[h]
  \includegraphics[width=3.0in]{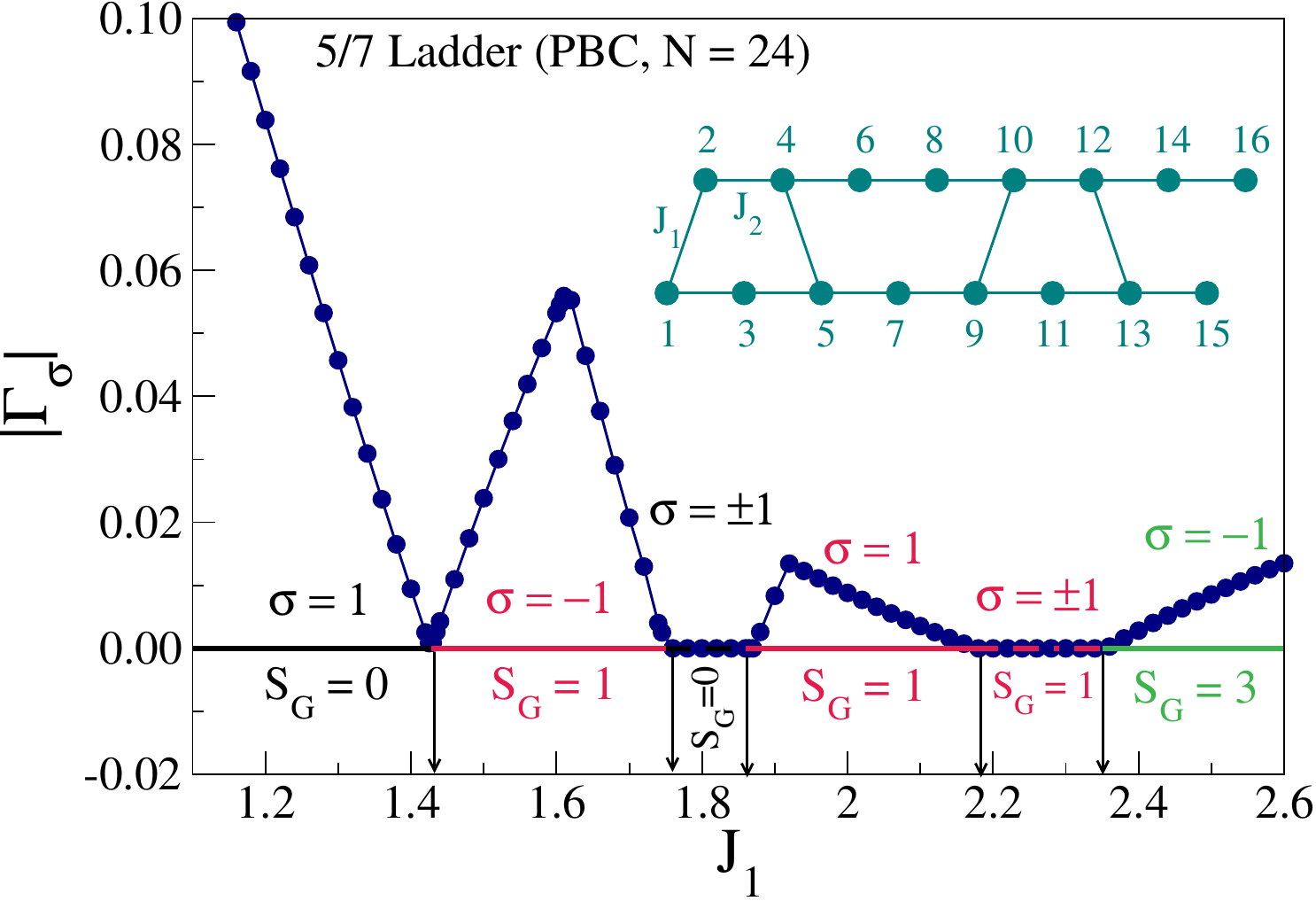}
  \caption{\label{fig_57_res3}Energy gap $|\Gamma_{\sigma}|$ between the lowest energy states belonging to two symmetry sectors $\sigma=\pm 1$ of a 5/7 ladder of 24 spins with PBC. The GS spin in the system is also shown. Reproduced from ref.~\cite{geet} with permission from APS, copyright 2023.}
\end{figure}
\begin{figure}[h!]
	\includegraphics[width=3.0 in]{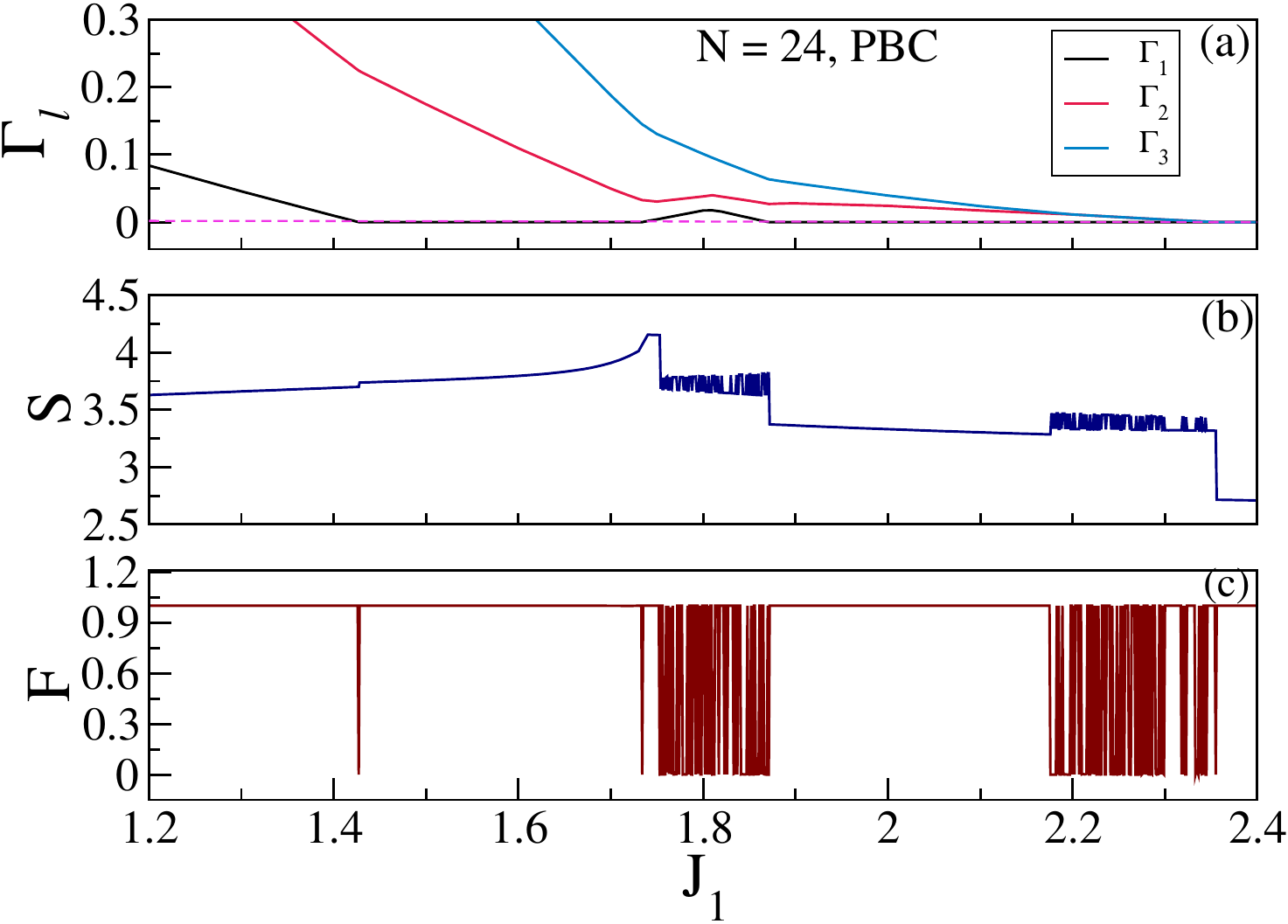}
	\caption{\label{fig:nosym_all_57}(a) The spin gaps $\Gamma_1$, $\Gamma_2$ and $\Gamma_3$ are shown as a function of $J_1$ for 24 site spins in a 5/7 skewed ladder with PBC. For $J_1 < 1.427$, $\Gamma_1$ becomes nonzero and the system shows nonmagnetic behaviour. The system enters a reentrant nonmagnetic phase for $1.734 < J_1 < 1.872$ where $\Gamma_1$ is nonzero. The EE for the unsymmetrized GS is shown in (b), and the fidelity is shown in (c) as a function of $J_1$ for the unsymmetrized GS. The EE exhibits a discontinuous change, while fidelity shows a sharp drop at the transition points. Wild fluctuations (many transitions) in both the EE and fidelity around $J_1=1.8$ and 2.3 are due to the degeneracy in the GS energy. Reproduced from ref.~\cite{Das} with permission from Springer Nature, copyright 2023.}
\end{figure}
The ladder also has a mirror plane of symmetry, $\sigma$, perpendicular to the plane of the ladder and passing through the points $(i,7)$  and the midpoint on the $(i,6)$ -- $(i,7)$ bond where $i$ is the unit cell index. The eigenstates thus belong to either even (e) or odd (o) subspace of the point group $(E, \sigma)$. When the lowest energy states in the two subspaces, `e' and `o' are degenerate, the $\sigma$ symmetry of the states is broken. This implies that the GS has a bond order wave. Similarly, the system has a spin inversion symmetry, $P$, corresponding to rotation of all the spins by $\pi$ about the y-axis. The eigenstates can therefore be classified as either `$+$' or `$-$' under the symmetry group $(E,P)$. In spin-1/2 systems, the eigenstates in the `$+$' subspace have a total spin which is odd, while in the `$-$' subspace the eigenstates have a total spin that is even. Again, if the lowest energy states of the two subspaces are degenerate, the spin inversion symmetry is broken resulting in a spin density wave. In this ladder system, in the singlet GS, we see for the finite system, breaking of the mirror symmetry for $1.75 < J_1 < 1.87$ (Fig.~\ref{fig_57_res3}).
Another interesting feature of the phase diagram is in the region $2.18 < J_1 < 2.35$ where in the triplet GS the mirror symmetry is broken. This leads to a spin current in the rings and is known as breaking of the vector chiral symmetry. What is unusual about the 5/7 skewed ladder is that this symmetry is broken in the absence of either an applied magnetic field or anisotropy in the spin exchange interactions. In Fig.~\ref{fig:nosym_all_57}, we show the entanglement entropy and fidelity of the 5/7 ladder as a function of the parameter $J_1$. We note that the transitions clearly show up as jumps in the fidelity and discontinuities in the entanglement entropy, wherever there are jumps in the spin gap.
\subsection{3/4 Skewed ladders}
\begin{figure}[h]
  \includegraphics[width=3.0in]{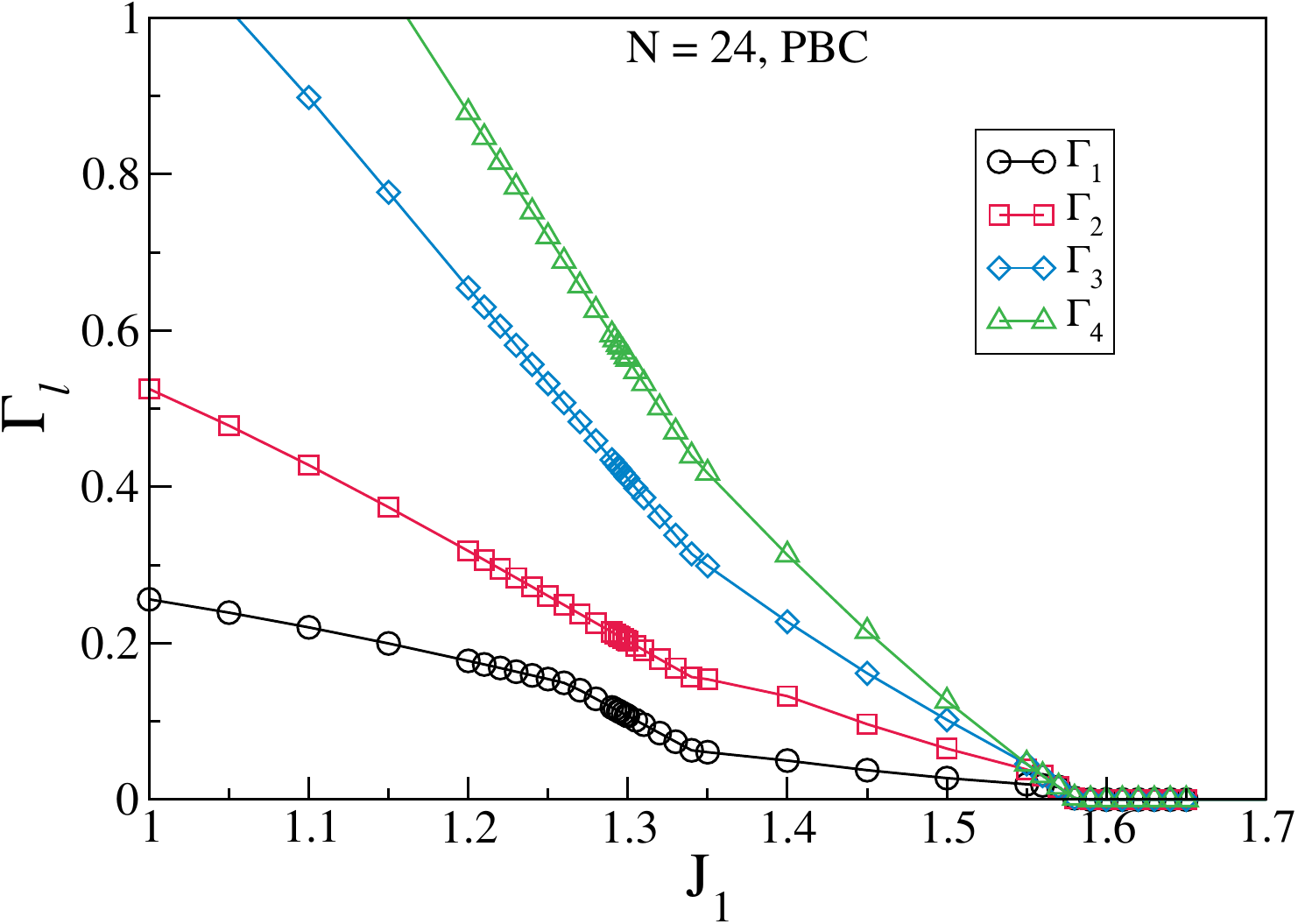}
	\caption{\label{fig_34_gap} Spin gaps $\Gamma_l$ for a 3/4 skewed ladder of 24 spins with PBC shown as a function of $J_1$. The GS spin $S_G$ = 0 for $J_1 \le 1.581$ and $S_G = 4$ for $J_1 \ge 1.581$. Reproduced from ref.~\cite{Das} with permission from Springer Nature, copyright 2023.}
\end{figure}
\begin{figure}[h]
  \includegraphics[width=3.0in]{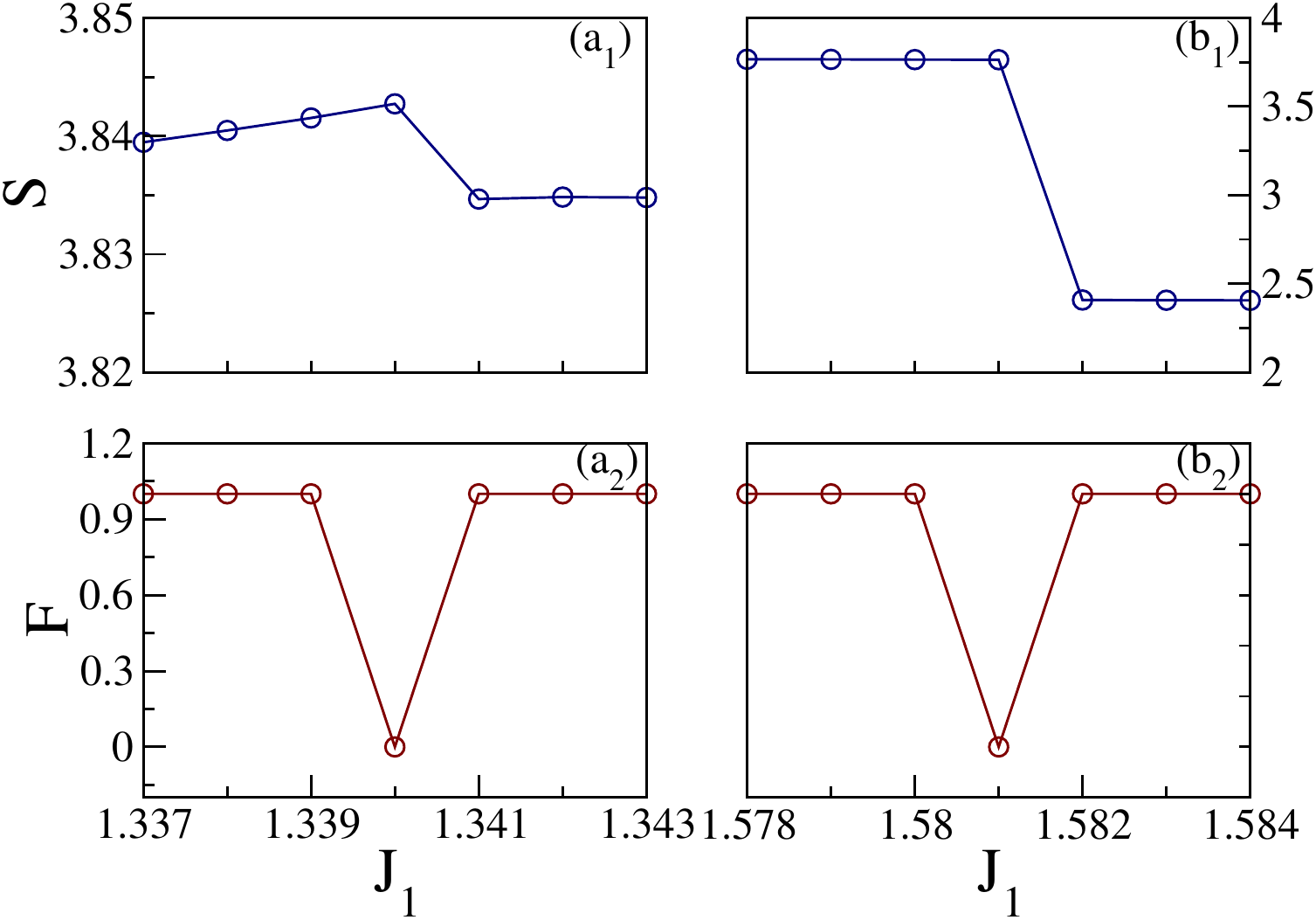}
	\caption{\label{fig_EE_fidelity_34} The behaviour of EE (S) and fidelity (F) in the range $1.337 \le J_1 \le 1.343$ of a 3/4 skewed ladder with N = 24 spins are shown in ($a_1$) and ($a_2$) respectively. The EE changes abruptly and a sharp drop in fidelity occurs at $J_1 = 1.340$. At other values of $J_1$ fidelity is constant (= 1). Simillary the behaviour of EE and fidelity in the range $1.578 \le J_1 \le 1.584$ are shown in ($b_1$) and ($b_2$). At $J_1 = 1.581$, the entropy changes and fidelity shows a sharp drop. Reproduced from ref.~\cite{Das} with permission from Springer Nature, copyright 2023.}
\end{figure}
These are perhaps the simplest of the skewed ladder systems that we have studied. The GS spin $S_G$ increases with increase in $J_1$ and saturates at $S_G = p/2$, where $p$ is the number of triangles in the ladder, for $J_1 \ge 1.581$ (Fig.~\ref{fig_34_gap}). The unit cell has six spins and the spin gaps as a function of $J_1$ for four unit cells is shown in Fig.~\ref{fig_34_gap}. For large $J_1 (> 5)$ the spin correlation between spins of the triangle along the rungs is antiparallel $(-0.5)$ and along the edge of the triangle lying on the leg of the ladder is $(+0.25)$. For very large $J_1$, in a triangle, the spin density at the apical sites is $-1/3$ while at the basal sites is $+2/3$. What is interesting is the fact that for ladders with open boundary condition,  GS attains the maximum possible spin of $S_G = -1/3 + N/6$, where $N$ is the number of sites in the ladder, for $J_1 \ge 1.59$ for all system sizes. The phase transitions also show up both in entanglement entropy and in fidelity studies (Fig.~\ref{fig_EE_fidelity_34})~\cite{Das}. The fidelity studies are helpful, in confirming to a high degree of accuracy, the phase boundaries.
\subsection{3/5 Skewed ladders}
\begin{figure}
\includegraphics[width=3.0 in]{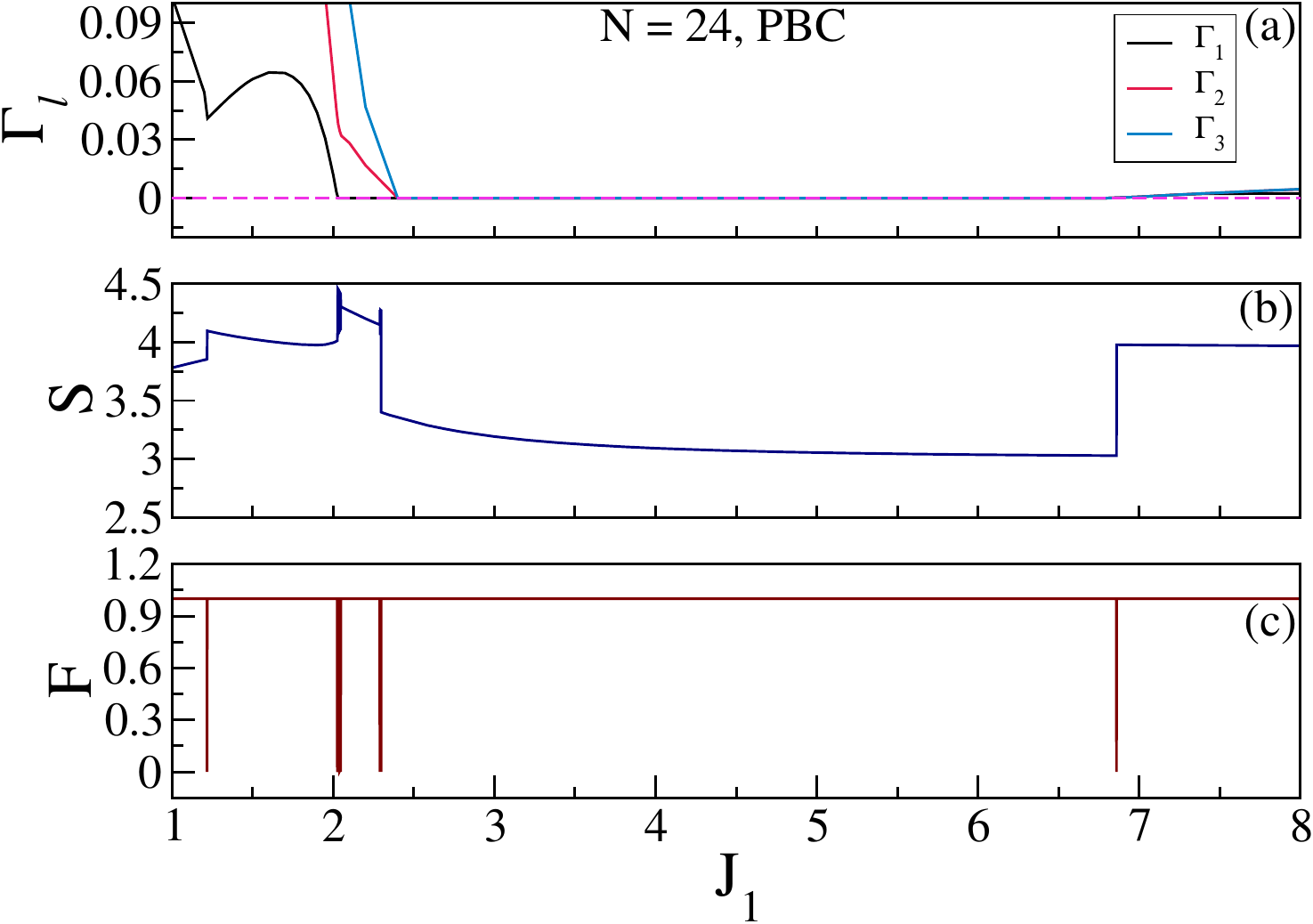}
	\caption{\label{fig:gap_35}(a) The spin gaps $\Gamma_1$, $\Gamma_2$ and $\Gamma_3$ are shown as a function of $J_1$ for N = 24 sites in a 3/5 skewed ladder with PBC. For $J_1$ $<$ 2.026 and $J_1$ $>$ 6.859, $\Gamma_1$ becomes 0 and the system shows nonmagnetic behaviour. In the region $2.026 \le J_1 \le 2.297$ the GS spin gradually changes from 0 to 3. (b)shows the EE and (c) shows the fidelity for the unsymmetrized GS as a function of $J_1$. The EE exhibits a discontinuous change, while fidelity shows a sharp drop at the transition points. The thick lines in the entanglement entropy and fidelity in the neighborhood of $J_1=2$ is due to a change in the reflection symmetry of the ground state of same spin. Reproduced from ref.~\cite{Das} with permission from Springer Nature, copyright 2023.}
\end{figure}

We have studied a 3/5 ladder with $24$ spins and periodic boundary conditions by the exact
diagonalization method. The finite size effects are weak since the spin-spin correlations decay very rapidly and studies on the 24 site system is close to the thermodynamic limit. The GS
\begin{figure}[h]
  \includegraphics[width=3.0in]{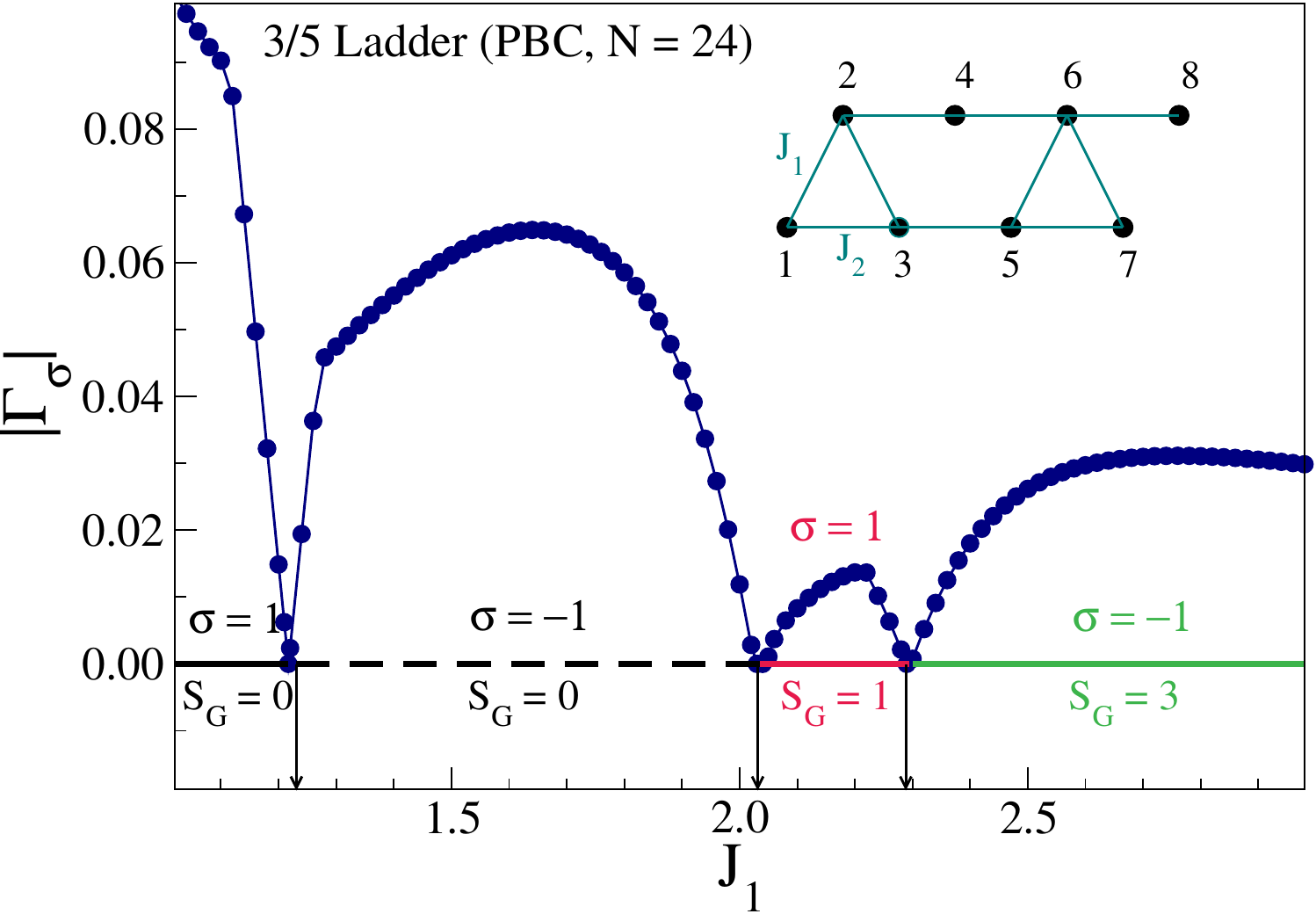}
  \caption{\label{fig_35_result1}Quantum phases in the GS of the 3/5 ladder with PBC and 24 spins; inset shows two unit cells. Reproduced from ref.~\cite{geet} with permission from APS, copyright 2023.}
\end{figure}
\begin{figure}[h]
  \includegraphics[width=3.0in]{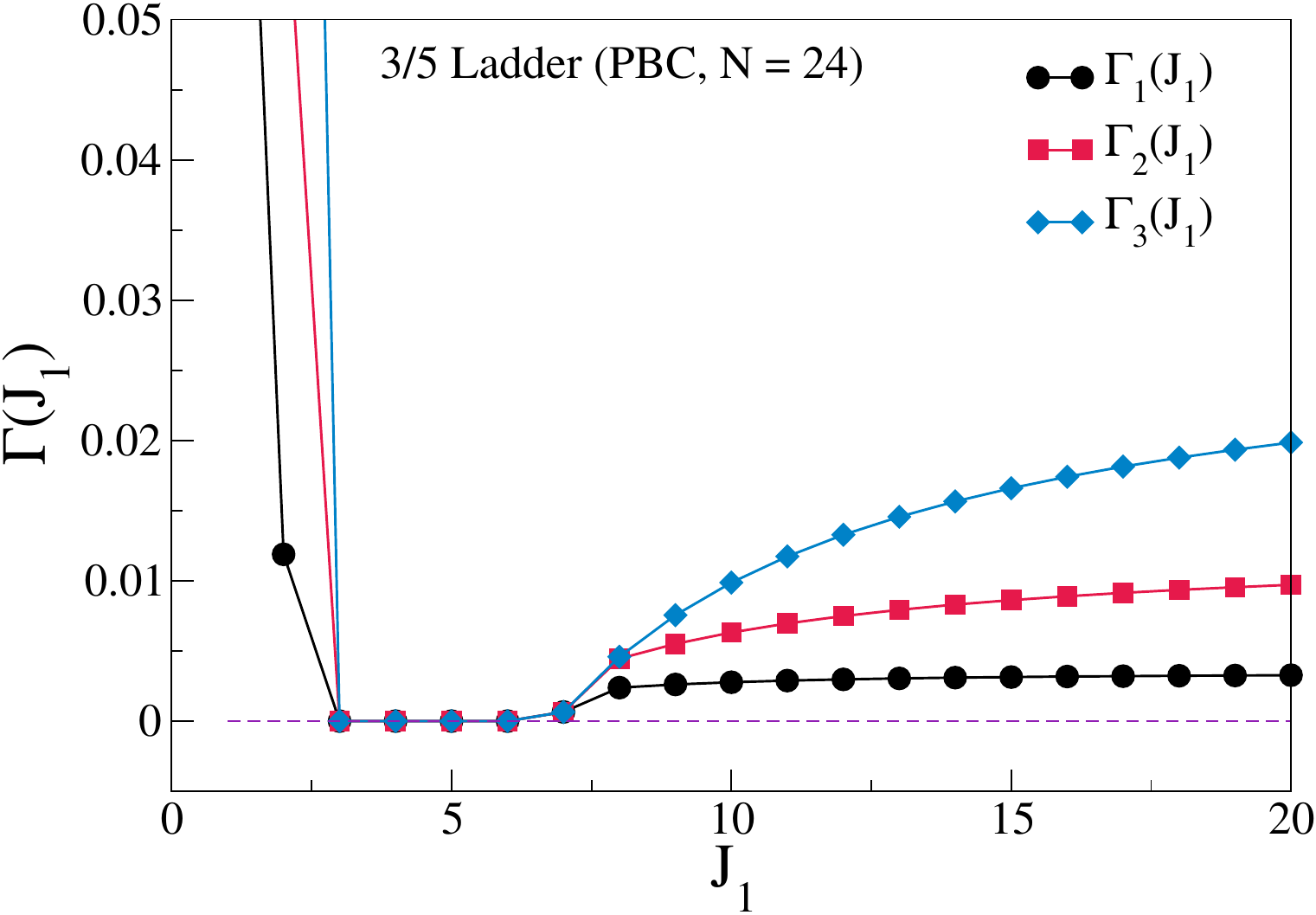}
	\caption{\label{fig_35_result2}Excitation energy gaps from the $S^z = 0$ sectors to higher $S^z$ sectors in 3/5 ladder with PBC and 24 spins. Reproduced from ref.~\cite{geet} with permission from APS, copyright 2023.}
\end{figure}
of this system is a singlet for $J_1$ up to $2.03$ where it switches to spin 1 and then to spin 3 for $J_1 = 2.3$. This system also has a mirror plane of symmetry and the singlet GS switches from even subspace to odd subspace at $J_1 = 1.22$. The GS symmetry remains odd until the system switches to a triplet GS at $J_1 = 2.03$ (Fig.~\ref{fig_35_result1}). In the triplet state, the symmetry is even until $J_1 = 2.30$ at which point the system switches to a spin 3 GS. In the spin 3 GS, the symmetry remains odd until $J_1 = 3$. Beyond this value of $J_1$, we have not followed the symmetry of the GS. The degeneracy at $J_1 = 1.22$ shows that the system will have a bond order wave instability. One  unusual feature of the 3/5 ladder is that the GS becomes nonmagnetic for $J_1$ values beyond 6.87 (Fig.~\ref{fig_35_result2}) and stays a singlet until the highest value of $J_1$ that we have studied. This behavior of a re-entrant singlet phase has never before been observed in spin ladders or indeed spin chains. Entanglement entropy and Fidelity variations are shown in Fig.\ref{fig:gap_35}. Using these studies we can clearly pin point these quantum phase transitions.
\section{\label{sec:plateau}Magnetization Plateaus}
\begin{figure}
  \includegraphics[width=3.2 in]{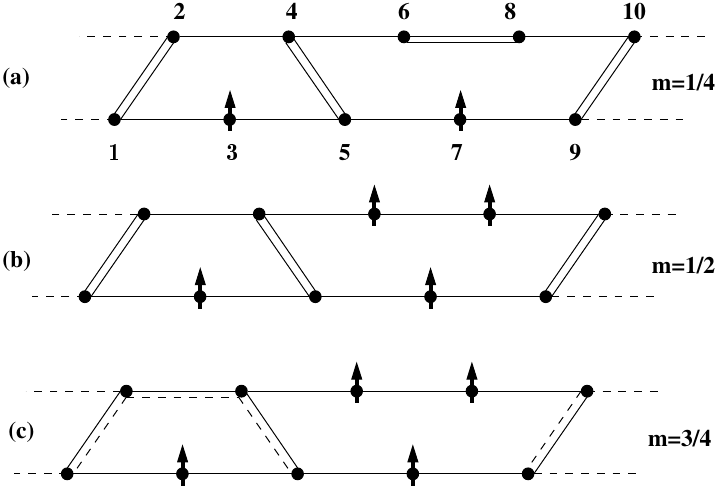}
  \caption{\label{plateau_config}Schematic representation of the spin configurations in (a) $m=1/4$, (b) $m = 1/2$ and (c) $m=3/4$ plateau phases. Hollow bond represents the singlet bond formation and  arrow represents a free spin aligned along z-direction. Resonating bonds in (c) represent quadramers in triplet state. Reproduced from ref.~\cite{plateau57} with permission from APS, copyright 2023.}
\end{figure}

In the previous section, we saw that GS properties of various types of skewed ladder structures exhibit exotic quantum phases upon tuning the exchange parameters. In this section, we explore the role of an external magnetic field $B$ on the GS properties of the ladders. In many frustrated systems, magnetization and magnetic field curves ($M-B$ curves) exhibit plateau like structures~\cite{kikuchi2005,kikuchi2006,gu2006,hase2006,zhao2010,hardy2004,wang2011,hardy2006,ishiwata2005,yao2012,lenertz2011,he2009} and  the maximum number of plateaus in the thermodynamic limit in a one dimensional system is given by OYA~\cite{oya97} condition which follows for a spin-$S$ system by generalizing the Lieb-Schultz-Mattis (LSM) theorem~\cite{affleck86,lsm61}. The OYA condition for observing a plateau at magnetization $m$, which is a fraction of the saturation magnetization, is given by $Sp(1-m) \in \mathbb{Z}$, where $S$ is the spin of a site, $p$ is the number of lattice sites per unit cell, and $\mathbb{Z}$ represents the set of positive integers. The condition is further generalized to an n-leg ladder, and is given as $nSp(1-m) \in \mathbb{Z}$~\cite{cabra97,cabra98}. In this section we discuss the quantum phases in 5/7 skewed ladder in the presence of an external magnetic field $B$, and find that the $m-B$ curve in this system also exhibits interesting plateau like structures. A plateau in the curve represents stable GS in the presence of a field which has a large excitation gap to the lowest excited state for values of $B$ in the range $B_{b} < B < B_{e}$ where $B_b$ and $B_e$ are the magnetic fields at the beginning and end of a plateau respectively. Before going to plateau phases, let us discuss the structure of the 5/7 skewed ladder. There are eight spins and ten bonds in a unit cell with two rung bonds having exchange interaction $J_1$ and other eight bonds along the leg having exchange strength $J_2$ as shown in Fig.~\ref{plateau_config}. In the absence of the magnetic field the GS can be tuned from a non-magnetic state to a high-spin GS where one quarter of polarized spins at sites 3 and 7 in each unit cell are ferromagnetically coupled to polarized spins of neighboring unit  cells in the large $J_1 \ (> 2.35 J_2)$ limit. The remaining six spins form three singlet dimers in each unit cell~\cite{plateau57}. Using the OYA condition One can expect plateaus at $m = 0$, 1/4, 1/2 and 3/4 for this system. In this section, we show that there are indeed three plateaus at $m = 1/4$, 1/2 and $3/4$ in this skewed ladder system  and the number of plateaus makes this system more interesting compared to the regular zigzag chain which has a lone plateau at $m = 1/3$, although the OYA condition predicts two more plateaus at $m = 0$ and 2/3~\cite{meisner2007, tandon99, okunishi_jpsj2003, okunishi_prb2003}. The 5/7 system is quite unique as it exhibits  at least three plateaus in two leg ladders. The Hamiltonian conserves $S^z$ even in the presence of an axial field, therefore field dependent GS energies are obtained simply by adding the Zeeman term to the zero  field energies,
\begin{equation}
  E(S^z, B) = E(S^z,B=0) - B S^z
\end{equation}
where $S^z$ is the z-component of the total spin and $E(S^z, B)$ and $E(S^z,B=0)$ are the  lowest energy states in the chosen spin sector with and without an applied magnetic field $B$. The change in magnetization from $S^z$ to $S^z + 1$ occurs at a finite field $B$ when $E(S^z, B)$ line intersects the $E(S^z+1, B)$ line. The magnetization of the system is the value of $S^z$ corresponding to lowest energy state at finite $B$.
\begin{figure}
  \includegraphics[width=3.2 in]{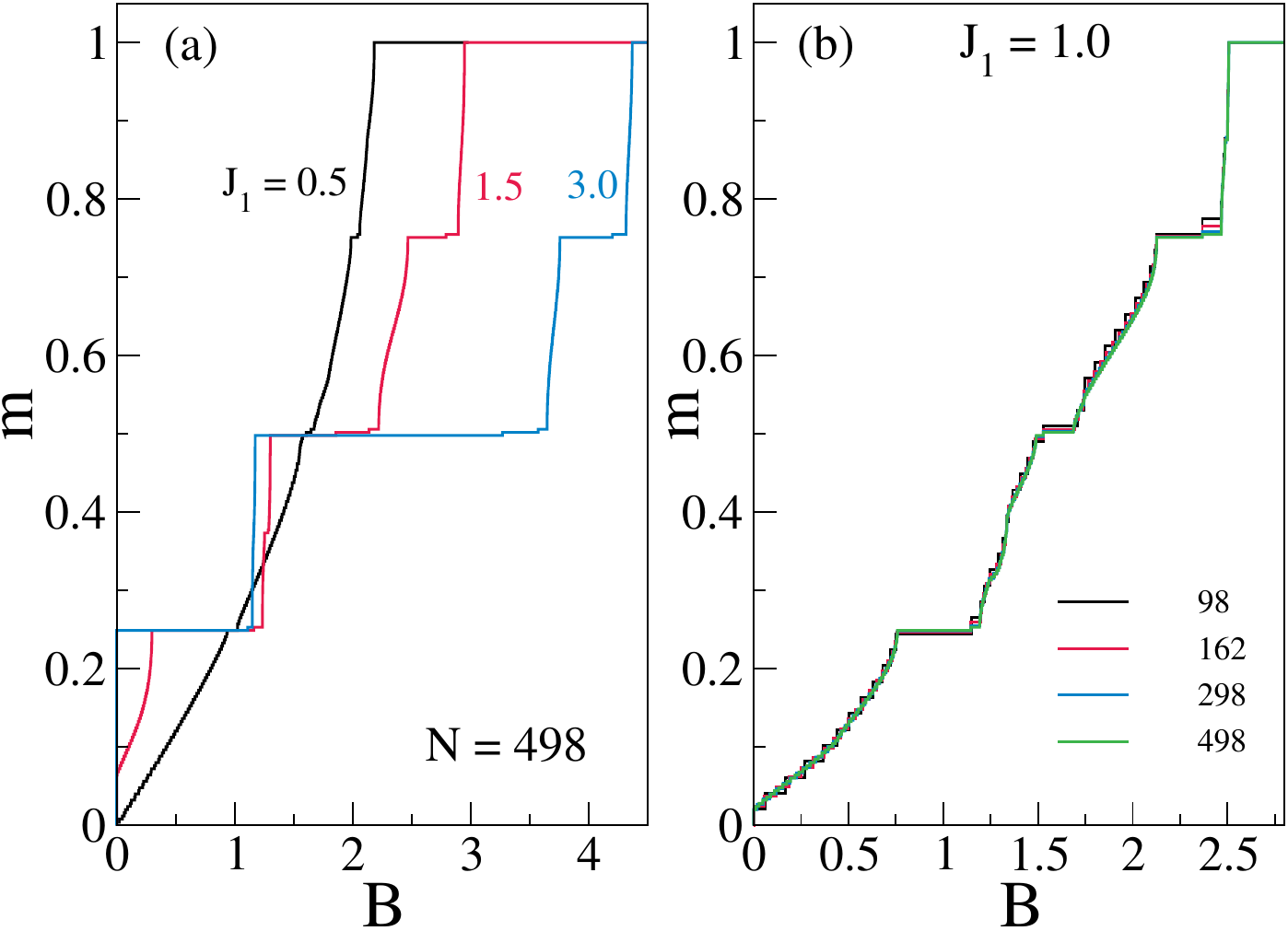}
  \caption{\label{plateau_mB}(a) The normalized magnetization as a function of external magnetic field $B$ for different $J_1$ for $N=498$. (b) The finite size effect of the magnetization plateaus for $J_1 = 1.0$, leading to unstable plateaus for intermediate $m$ values. Reproduced from ref.~\cite{plateau57} with permission from APS, copyright 2023.}
\end{figure}

The $m-B$  curves of 5/7 system for different $J_1$ values in units of $J_2$ are shown in Fig.~\ref{plateau_mB}.  Width of plateaus are  noticeable only for  $J_1 >0.5$ and  in small $J_1$ limit a finite $B$ is required to reach the first plateau at $m=1/4$. The finite size effect on the plateau width is negligible as shown in the Fig.~\ref{plateau_mB}(b). However, for large $J_1 >2.35$ the GS has $m=1/4$ even in the absence of the field $(B = 0)$. In this phase the rung singlets are strong, and an effective ferromagnetic interaction develops between spins 3 and 7  of the unit cell as shown in Fig.~\ref{plateau_config}(a). In fact most of these plateaus in the 5/7 ladder are formed because of the strong dimer formations in the system.

\begin{figure}
  \includegraphics[width=2.2 in]{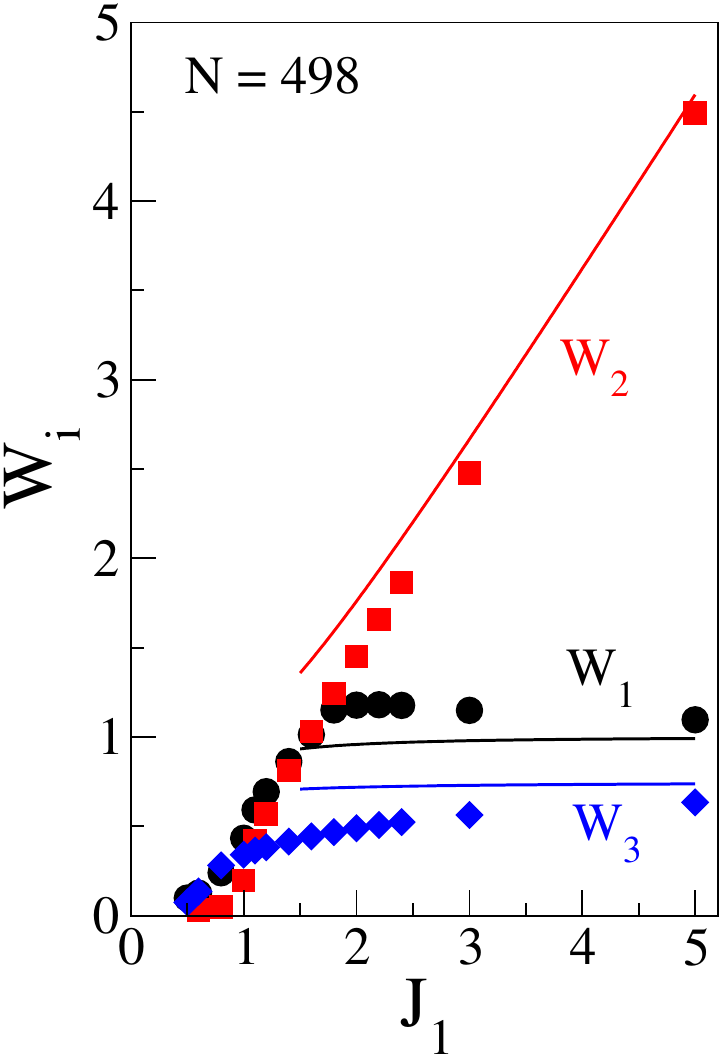}
  \caption{\label{fig:pltwd} The magnetization plateaus $W_1, \ W_2, \ W_3$ are plotted as a function of $J_1$ and the solid lines are obtained from the second order perturbation theory calculation in the large $J_1$ approximation. Reproduced from ref.~\cite{plateau57} with permission from APS, copyright 2023.}
\end{figure}

In small $J_1$ limit the effective exchange between 3 and 7 is still antiferromagnetic and the GS is non-magnetic. Therefore, a finite $B=B_{1/4}$ is required to achieve $m=1/4$ plateau and  $B_{1/4}$ decreases with increasing $J_1$ as shown inFig.~\ref{fig:pltwd}. All possible plateaus for $J_1 >0.5$ have finite width for rung exchange $J_1 >0.5$. For $J_1>2.35$, $m=1/4$ plateau is a GS for $B=0$. The plateau widths of 1/4 ($W_1$) and 3/4 ($W_3$) weakly depend on $J_1$ whereas
the width of the 1/2 plateau ($W_2$) is proportional to $J_1$ as shown in Fig.~\ref{fig:pltwd}.
\begin{figure}
  \includegraphics[width=3.3 in]{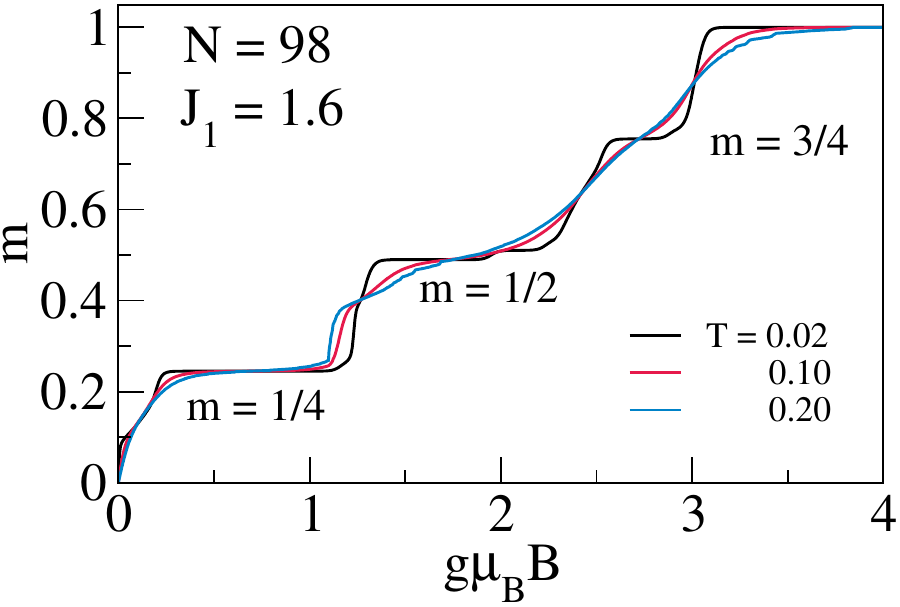}
  \caption{\label{fig:plt_T} $m-B$ curve at different temperatures. Reproduced from ref.~\cite{plateau57} with permission from APS, copyright 2023.}
\end{figure}

The thermal stability of magnetic plateau is another important aspect for usefulness of these systems for device fabrication. In Fig.~\ref{fig:plt_T} the stability of these plateaus are shown for three temperatures in reduced units $J_2/k_B = 0.02$, 0.1 and 0.2 for $J_1=1.6$. We notice that only $m=1/4$ and 1/2 survives any finite thermal fluctuations. The  $m=3/4$ plateau shows significant fluctuation for any finite $B$ and this plateau has the smallest width out of all three plateaus. The dominant fluctuations in 3/4 plateau may be due to small magnetic gap about the GS at that field value. The gap around the $m=1/4$ and 1/2 GS is large and it suppresses the low temperature thermal fluctuation.
\section{\label{sec:summary}Summary and outlook}
In this review we summarized the GS properties of skewed ladders both in presence and absence of a magnetic field. The skewed ladders can be viewed as periodically missing rung bonds in the zigzag ladders. We began with a brief discussion of the quantum phase diagram of a zigzag ladder or $J_1-J_2$ chain model and effect of dimerization is also discussed based on reference~\cite{chitra95,Pati1997}. The GS properties of the 5/7, 3/4, and 3/5 skewed ladders are discussed in detail based on the earlier work~\cite{geet,plateau57,Das}.  For 5/7 ladder and for fixed system size, while $S_G$ increases with $J_1$ till $J_1$ = 1.75, thereafter, the system reenters the singlet GS for 1.75 $<$ $J_1$ $<$ 2.18. The system rapidly regains higher $S_G$ values for $J_1 = 2.18$  and attains the highest GS spin of $S_{G} = n$. In the re-entrant phase (1.75 $<$ $J_1$ $<$ 2.18), the spin correlations indicate a long wavelength spin density wave. It is also noted that this system has unusual symmetry breaking in the absence of either an applied magnetic field or anisotropy in the spin exchange interactions and the quantum phase boundary can be determined using the entanglement entropy etc ~\cite{Das}. In 3/4 ladder the GS spin $S_G$ increases with increase in $J_1$ and saturates at $S_G = p/2$, where $p$ is the number of triangles in the ladder, for $J_1 \geq 1.581$ as shown in (Fig.~\ref{fig_34_gap}). The GS of 3/5 system switches back and forth from non-magnetic to magnetic states and the GS is doubly degenerate at various parameter regimes. The GS of these skewed ladders are very different from the zigzag~\cite{hase2004, masuda2004, mizuno98, riera95, kamieniarz2002, drechsler2007, dutton2012, dd2018, sirker2010} and normal two leg ladder~\cite{Sandvik1,Azuma_1994,Dagotto,Dagotto_review} where the GS is always a singlet.

The magnetization plateaus in these systems show interesting features, when an axial magnetic field is applied. We have also studied the GS properties of the 5/7-skewed ladder which show three magnetic plateaus at $m =$ 1/4, 1/2 and 3/4 and these plateaus conform to the OYA criterion~\cite{oya97}. The magnetic plateaus of these systems are very different from zigzag and normal ladder which have a single plateau at $m$=1/3~\cite{okunishi_jpsj2003} and 0~\cite{oya97,Hida_m_0} respectively. While we have presented a theoretical construct of these systems, we believe that these systems can be achieved experimentally in concatenated inorganic transition metal complexes. The 5/7 skewed ladder system can be designed as a molecular system having fused five- and seven-membered carbon rings~\cite{thomas2012}. Such a system corresponds to a fused azulene lattice~\cite{thomas2012}, and this type of structures may be found at the grain boundary of a graphene sheet~\cite{Huang2011,Kochat,Balasubramanian2019}. The 3/4 skewed ladders can be mapped into a connected trimers system and azurites are some possible candidates~\cite{Kang_azurite,Honecker_azurite,Rule_azurite}. Designing magnetic skewed ladder systems is one of the important goal for future studies for these systems. These systems have interesting topological properties in the GS and further study is required to understand the relation between the topological and magnetic structure in these systems.
\section*{Conflicts of interest}
There are no conflicts of interest to declare.
\section*{Acknowledgements}
SR thanks Indian National Science Academy for financial support. MK and RR thank DST-SERB for funding through projects CRG/2020/000754 and CRG/2019/001627 respectively.

\bibliography{reference} 

\end{document}